\documentclass[
preprint,
superscriptaddress,
amsmath,
amssymb,
showkeys,
aps,
prb,
]{revtex4-2}

\usepackage{threeparttable}
\usepackage{floatrow}
\usepackage{setspace}
\usepackage{makecell}
\usepackage{amsmath}
\usepackage{graphicx}
\usepackage{dcolumn}
\usepackage{bm}
\usepackage{hyperref}
\usepackage[mathlines]{lineno}
\usepackage{textcomp}
\usepackage{gensymb}
\usepackage{subfigure}
\usepackage{url}
\usepackage{xcolor}
\usepackage{hhline}
\usepackage[version=3]{mhchem}
\usepackage{multirow}
\usepackage[american]{babel}
\usepackage{hhline}
\usepackage{textgreek}
\usepackage{miller}
\usepackage{lipsum}
\definecolor{red}{rgb}{0.0,0.0,0.0}
\definecolor{green}{rgb}{0.1,0.6,0.1}
\definecolor{yellow}{rgb}{0.8,0.8,0.6}
\definecolor{blue}{rgb}{0.1,0.1,0.6}
\begin{document}

% \title{Anisotropic Kinetics of Ion-Irradiation-Induced Phase Transition in Gallium Oxide}

\title{Anisotropic Kinetics of Ion-Irradiation-Induced Phase Transition in Gallium Oxide}

\author{Taiqiao Liu}
\affiliation{School of Integrated Circuits, Wuhan University, Wuhan, 430072, China}

\author{Tongtong Wang}
\affiliation{School of Integrated Circuits, Wuhan University, Wuhan, 430072, China}

\author{Zeyuan Li}
\affiliation{School of Power and Mechanical Engineering, Wuhan University, Wuhan, 430072, China}

\author{E Zhou}
\affiliation{School of Integrated Circuits, Wuhan University, Wuhan, 430072, China}

\author{Junlei Zhao} 
\email{junlei.zhao@mrdi.org.hk}
\affiliation{The Hong Kong Microelectronics Research and Development Institute, Hong Kong, 999077, China}

\author{Jiaren Feng}
\affiliation{School of Integrated Circuits, Wuhan University, Wuhan, 430072, China}

\author{Xiaoyu Fei}
\affiliation{School of Integrated Circuits, Wuhan University, Wuhan, 430072, China}

\author{Yuzheng Guo}
\affiliation{School of Power and Mechanical Engineering, Wuhan University, Wuhan, 430072, China}

\author{Flyura Djurabekova}
\affiliation{Department of Physics and Helsinki Institute of Physics, University of Helsinki, P.O. Box 43, Helsinki, FI-00014, Finland}

\author{Sheng Liu}
\affiliation{School of Integrated Circuits, Wuhan University, Wuhan, 430072, China}

\author{Zhaofu Zhang}
\email{zhaofuzhang@whu.edu.cn}
\affiliation{School of Integrated Circuits, Wuhan University, Wuhan, 430072, China}

\begin{abstract}

Radiation-tolerant semiconductors have traditionally been engineered by the principle of suppressing defect accumulation and amorphization, based on the assumption that radiation damage is inherently stochastic.
Here we show that, in monoclinic $\beta$-\ce{Ga2O3}, a promising ultrawide-bandgap semiconductor, surface crystallographic orientation deterministically governs radiation tolerance through highly anisotropic kinetics of the $\beta$-to-$\gamma$ phase transition.
Using machine-learning molecular dynamics coupled with a local configurational-entropy descriptor, we quantitatively map anisotropic $\beta$-to-$\gamma$ transition kinetics, showing that the critical dose, transition-layer depth, and kinetic stability of the $\gamma$-phase are fundamentally governed by surface orientation.
Under ion irradiation, non-channeling surfaces such as \hkl(100), \hkl(001), and \hkl(-201) undergo severe surface amorphization, whereas the strongly channeling \hkl(010) surface resists damage accumulation and promotes subsurface $\gamma$-phase nucleation.
During thermal annealing recovery process, these initial states follow two distinct recovery pathways: the channeling \hkl(010) surface reverts directly from $\gamma$-to-$\beta$, whereas non-channeling surfaces follow a sequential amorphous-to-$\gamma$-to-$\beta$ transition pathway.
This work establishes surface orientation as a fundamental design principle for achieving radiation tolerance through controlled polymorphic transitions, providing a universal framework for engineering functional materials capable of withstanding extreme irradiation environments.

% We establish a robust local entropy benchmark for distinguishing between the amorphous, $\gamma$, and $\beta$ phases in \ce{Ga2O3}.

\end{abstract}

\maketitle
\subsection*{Introduction}

The reliability and lifetime of semiconductor devices in extreme radiation environments, such as outer-space exploration, aerospace, and nuclear energy, are critically limited by high-energy particle irradiation~\cite{Science, 2013Acta, 2023NMR}.
This leads to the accumulation of high-dose damage, which can induce significant device degradation originating from atomic-level processes, including atomic displacements, amorphization, and phase transitions, thereby ultimately alter the microscopic structure and macroscopic properties of materials~\cite{2018nordlund_primary, Weiguo_2019, he2024threshold}.
The search for radiation-tolerant materials capable of withstanding such extreme conditions has driven significant interest in ultrawide-bandgap semiconductors, among which gallium oxide (\ce{Ga2O3}) has emerged as a particularly promising candidate~\cite{2023radiation, 2025VanderWaalsβ-Ga2O3}.
Its appeal lies not only in its excellent electronic properties but also in its rich polymorphic landscape, which includes the stable $\beta$-phase and several metastable phases such as the $\kappa$-, $\alpha$-, $\delta$- and $\gamma$-phases~\cite{2020kbetagamma, gammaAM, 2025ZhangxiePhysRevLett}.

\begin{figure*}[!ht]
    \centering
    \includegraphics[width=15cm]{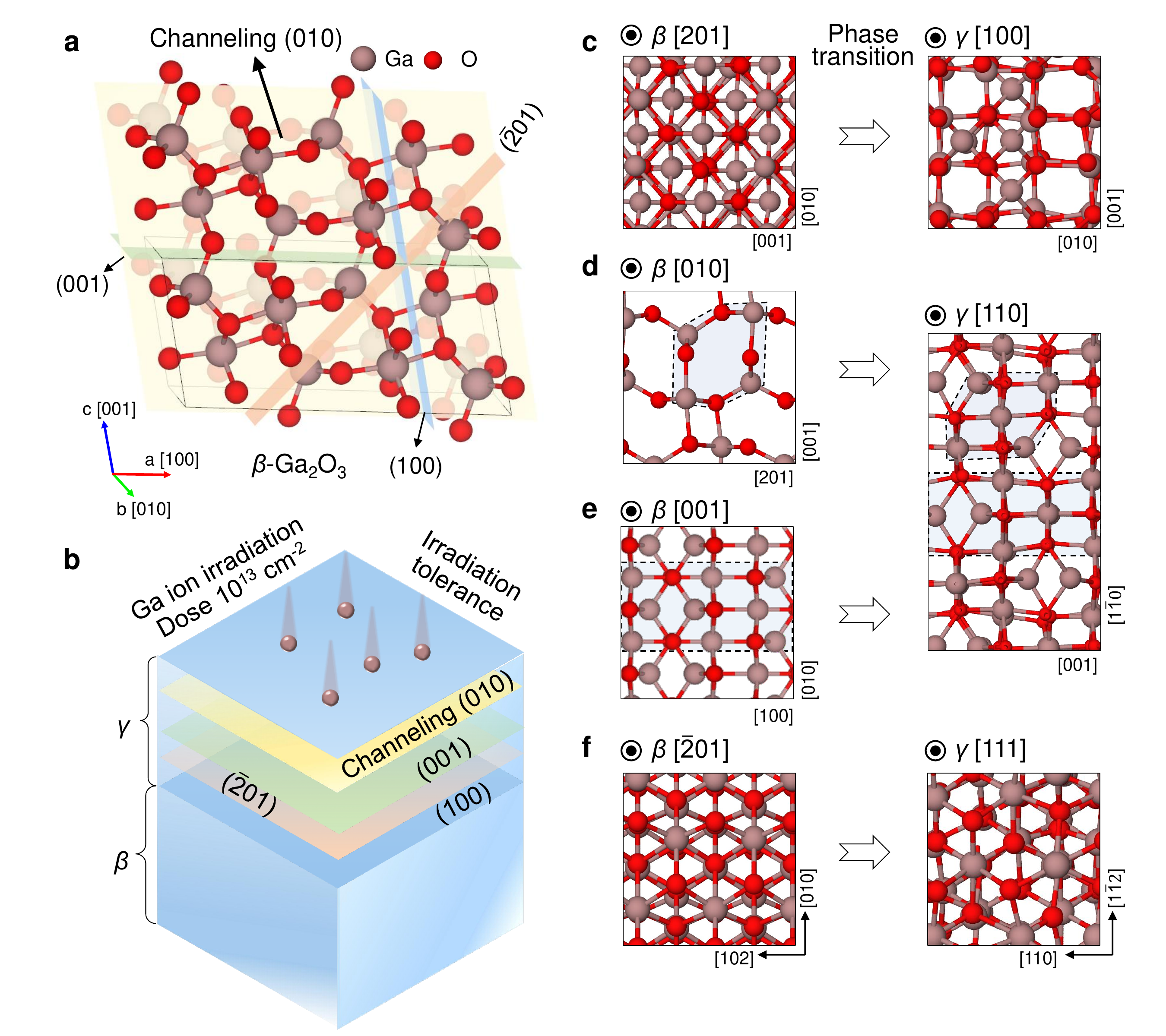}
    \caption{\textbf{Schematic of the anisotropic structure and irradiation-induced $\beta$-to-$\gamma$ phase transition in \ce{Ga2O3}.} 
    \textbf{a} Crystallographic planes of monoclinic $\beta$-\ce{Ga2O3} cleaved along the \hkl(100), \hkl(010), \hkl(001), and \hkl(-201) orientations. The \hkl(010) surface exhibits distinct ion channeling characteristics.
    \textbf{b} Illustration of the ion irradiation and the resulting $\gamma$-phase transition-layer depth after 4~ns annealing (as discussed below).
    \textbf{c-f} Orientation relationships between selected $\beta$-phase surfaces and the resulting $\gamma$-phase structures formed under irradiation: (c) $\beta$-\hkl(100) (viewed along $\beta$-\hkl[201]) to $\gamma$-\hkl(100), (d-e) $\beta$-\hkl(010)/\hkl(001) to $\gamma$-\hkl(110), and (f) $\beta$-\hkl(-201) to $\gamma$-\hkl(111) (equivalent to $\gamma$-\hkl(1-1-1)). 
    % {\red Angular deviation (1$\sim$2$^\circ$)} between the \hkl[-201] orientation and the normal to the \hkl(-201) plane in monoclinic $\beta$-\ce{Ga2O3}. 
    % The orientation defined by \hkl[110] and \hkl[1-12] has a normal direction of \hkl[111] (crystallographically equivalent to the calculated \hkl[1-1-1] in the cubic $\gamma$-\ce{Ga2O3}).
    }
    \label{fig:fig1}
\end{figure*}

Previous researches have demonstrated that irradiation can induce a $\beta$-to-$\gamma$ phase transition, resulting in a $\beta$/$\gamma$ dual-phase structure that substantially improves its radiation tolerance~\cite{2020kbetagamma, 2022APLgamma, 2023WangNanoLett, 2024APLmater, 2023radiation, 2023SiDopedPhaseTransition, 2024polymorphdiodes, 2025NanoLett, 2025PRMphasetransition, 2025PhysRevLett}.
\ce{Ga2O3} can effectively resist severe amorphization via $\beta$-to-$\gamma$ phase transition, even under irradiation doses as high as several hundred dpa (displacements per atom) at room temperature~\cite{2023radiation, 2025alphaNC, 2025iontrack}.
However, a critical disconnection persists between the mechanistic understanding of the $\beta$-to-$\gamma$ phase transition and the practical development of radiation-resistant \ce{Ga2O3}-based devices, whose electrical performance is governed by the $\beta$-\ce{Ga2O3} surface properties~\cite{LXHAPL2025}.
Existing investigations on the $\beta$-to-$\gamma$ phase transition have focused on specific crystallographic orientations, notably \hkl(010)~\cite{2023radiation, 2025NanoLett, 2025PhysRevLett, 2025scripta, EpitaxialRecovery2025} and \hkl(-201)~\cite{bektas2025defect}.
The distinct atomic configurations of the major \hkl(100), \hkl(010), \hkl(001), and \hkl(-201) planes in $\beta$-\ce{Ga2O3} (Fig.~\ref{fig:fig1}a) determine their anisotropic response to particle irradiation.
We have previously established that this anisotropy leads to fundamentally divergent primary damage mechanisms between non-channeling planes such as \hkl(100), \hkl(001), and \hkl(-201) and the channeling \hkl(010) plane~\cite{2025ActaTaiqiao}.
However, whether this initial disparity in damage logically leads to distinct phase transition pathways and, hence, the orientation-dependent kinetics of the high-dose-irradiation-induced $\beta$-to-$\gamma$ phase transition is yet to be explored. 

Thus, in the present work, we quantitatively resolve the anisotropic phase transition kinetics of $\beta$-\ce{Ga2O3} under ion irradiation (Ga$^+$, 5 keV) with a dose of $10^{13}$~cm$^{-2}$ at the atomic scale using machine-learning molecular dynamics (ML-MD) simulations (Fig.~\ref{fig:fig1}b).
By introducing the local configurational-entropy as a robust structural descriptor, we pioneeringly correlate the initial disorder with the final phase pathway, revealing that surface structure sets the stage for anisotropic damage evolution and ultimate phase stability.
We reveal that the $\beta$-to-$\gamma$ transition is a near-surface event whose critical dose and phase transition depth are not intrinsic to the bulk crystal but are decisively governed by surface orientation.
This work reveals a surface-governed, phase-transition-mediated radiation resistance mechanism in $\beta$-\ce{Ga2O3}, establishing a new design paradigm for oxide electronics operating in extreme irradiation environments.

% As shown in Fig.~\ref{fig:fig1}c-f, the $\beta$-\hkl(100) surface transformed into a $\gamma$-phase oriented along the \hkl(100) direction, exhibiting a honeycomb-like structural motif.
% In contrast, the $\beta$-\hkl(010) and \hkl(001) surfaces generated a $\gamma$-phase with a \hkl(110) crystallographic orientation, while the $\beta$-\hkl(-201) surface preferentially yielded a $\gamma$-phase oriented along the \hkl(111) direction.

\subsection*{Results and discussion}

As illustrated in Fig.~\ref{fig:fig1}c-f, the crystal structures of $\beta$- and $\gamma$-\ce{Ga2O3} differ primarily in their Ga sublattices.
The $\beta$-\hkl(100) surface (viewed along $\beta$-\hkl[201]) transforms into a $\gamma$-\hkl(100)-oriented surface, exhibiting a square-like structural motif.
In contrast, the $\beta$-\hkl(010) and $\beta$-\hkl(001) surfaces both generate a $\gamma$-phase with \hkl(110) orientation, whereas the $\beta$-\hkl(-201) surface preferentially yields a $\gamma$-phase aligned along the \hkl(111) direction.
Throughout the $\beta$-to-$\gamma$ phase transition, the O sublattices largely retains a face-centered cubic (FCC) stacking, with only minor local displacements and lattice distortion that reflect the symmetry change from monoclinic to cubic.
The transition is thus mediated by the radiation-induced migration of Ga atoms into new coordination environments, a kinetically favorable process supported by the concurrent recrystallization of the O sublattice~\cite{2023radiation, 2025PhysRevLett}.
Representative snapshots of the irradiation-induced $\gamma$-phase transition on the four $\beta$ surfaces are provided in Supplementary Figures~1 and 2.

Fig.~\ref{fig:fig2} presents a depth-resolved coordination analysis performed at 15-\r A intervals to probe the depth of the anisotropic phase transition layer.
Phase transitions are identified using Ga-Ga partial radial distribution function (PRDF) analysis, a proven method for differentiating the $\beta$- and $\gamma$-polymorphs~\cite{2023radiation, 2025PhysRevLett}.
As shown in Fig.~\ref{fig:fig2}b, the peak in the 4.0--5.2~\r A range corresponding to the second Ga-Ga coordination shell evolves to match that of $\beta$-\ce{Ga2O3} with increasing depth, providing strong evidence of a $\beta$-to-$\gamma$ phase transition near the surface (see Supplementary Figures~3-6 for full RDF/PRDF evolution).
To quantify this transition, we calculated Pearson correlation coefficients (Pr) based on the Ga-Ga PRDF profiles in the 4.0--5.2~\r A range~\cite{2025PhysRevLett}.
The critical phase transition depth for each surface was quantitatively defined as the point where the $\gamma$-phase and $\beta$-phase similarity trends intersect (Fig.~\ref{fig:fig2}c), with the critical depths of 66/65/60/75~\r A for \hkl(100)/\hkl(010)/\hkl(001)/\hkl(-201) surfaces, respectively, revealing a clear anisotropy in phase transition occurrence.
The dependence of RDF, PRDF and Pr coefficient on the irradiation dose is provided in Supplementary Figures~7--11.

\begin{figure*}[!ht]
    \centering
    \includegraphics[width=16cm]{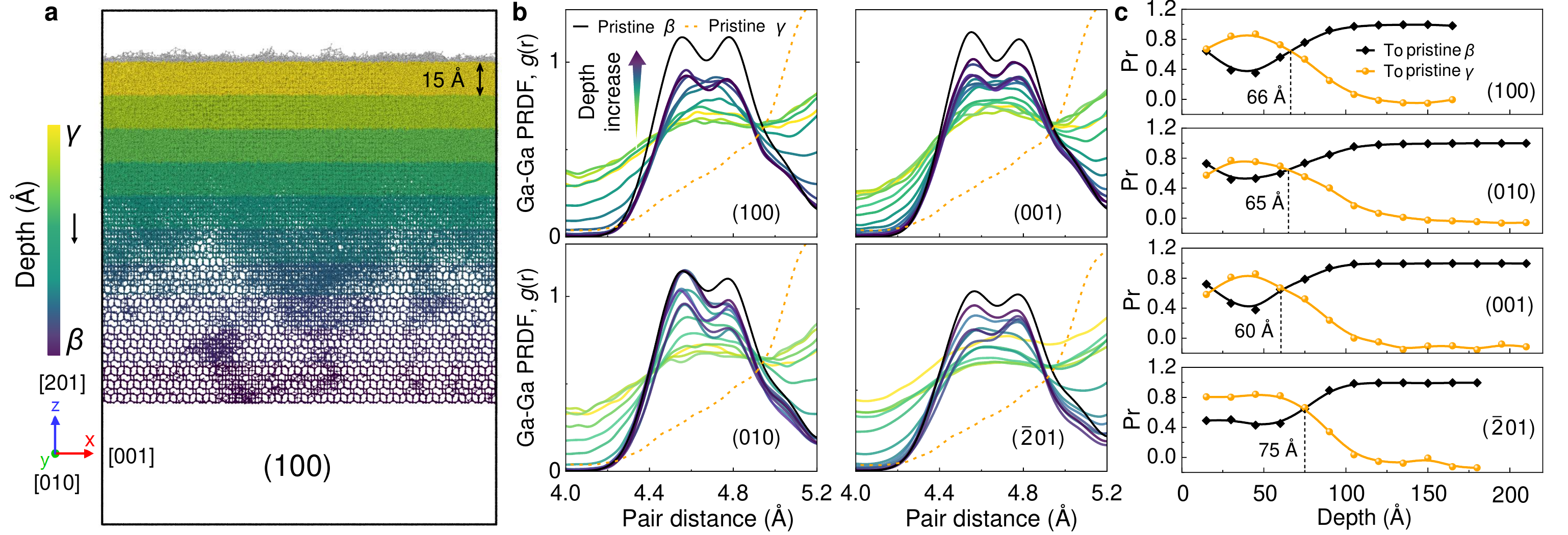}
    \caption{\textbf{Depth-resolved phase transition analysis.} 
    \textbf{a} Layer-colored cross-sectional model of irradiated $\beta$-\ce{Ga2O3}, illustrated using the \hkl(100) surface as an example, with layers sampled at 15~\r A intervals. 
    Surface atoms in the gray region (5~\r A) are not considered. 
    \textbf{b} Evolution of the Ga-Ga PRDF as a function of depth for the four surfaces. 
    \textbf{c} Depth-dependent Pr correlation coefficient representing structural similarity to pristine $\beta$- and $\gamma$-\ce{Ga2O3} for each surface. Total model depths of 150~\r A, 210~\r A, 210~\r A, and 180~\r A were considered for \hkl(100), \hkl(010), \hkl(001), and \hkl(-201) facets, respectively.
    }
    \label{fig:fig2}
\end{figure*}

\begin{figure}[ht!]
    \centering
    \includegraphics[width=16cm]{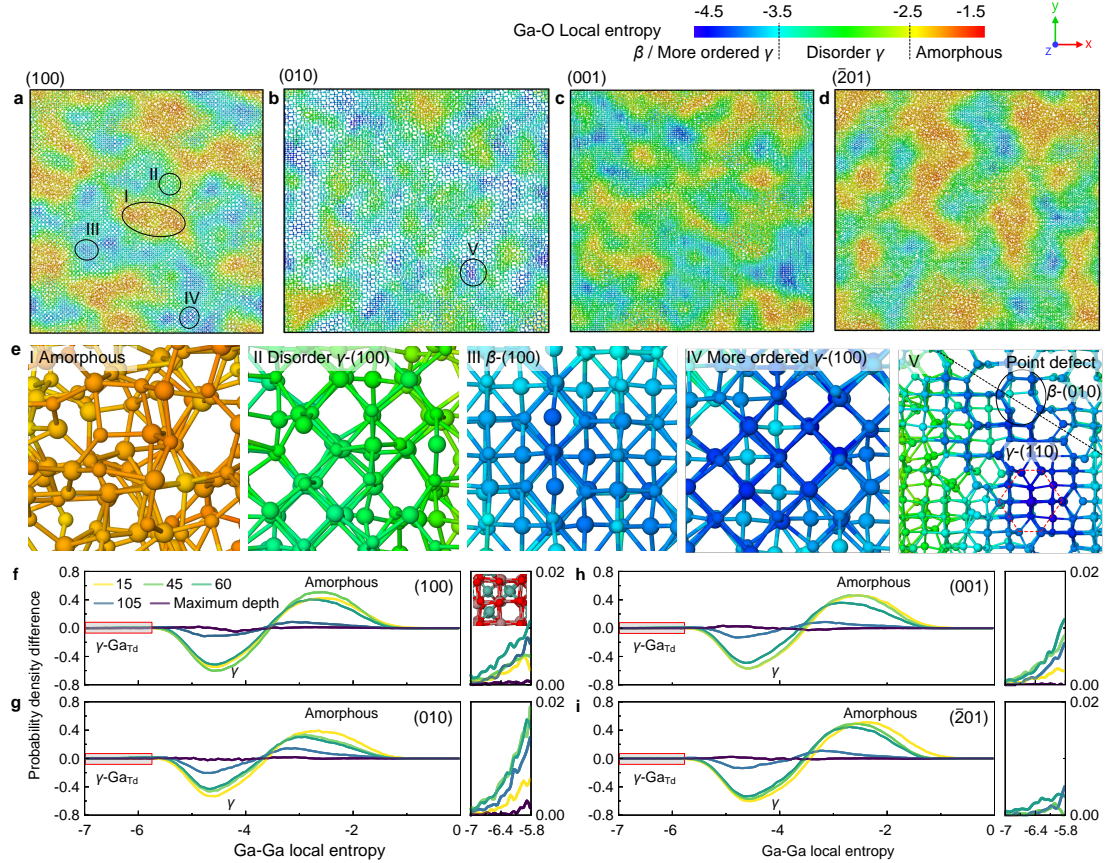}
    \caption{\textbf{Local entropy as a robust benchmark for identifying phase transition in disorder \ce{Ga2O3}.} 
    \textbf{a-d} Atomic structure at 60~\r A depth after irradiation for the four surfaces \hkl(100), \hkl(010), \hkl(001) and \hkl(-201), color-coded by Ga-O local entropy, with the associated scale defining the $\beta$-, disorder $\gamma$-, more ordered $\gamma$- and amorphous phases. 
    \textbf{e} Enlarged views of regions labeled from I to V in \textbf{e}: (I) amorphous structure, (II) disorder $\gamma$-\hkl(100), (III) $\beta$-\hkl(100), (IV) more ordered $\gamma$-\hkl(100) and (V) a mixed region of $\beta$-\hkl(010) and $\gamma$-\hkl(110). The red hexagonal markers are consistent with the experimental data from Ref.~\cite{gamma110APLMater}.
    \textbf{f-i} Radiation-induced changes in the probability density distribution of $\beta$-\ce{Ga2O3} for the four considered surfaces based on the Ga-Ga local entropy (see Supplementary Figure~21 for complete distributions). The magnified region outlined in red (range: $-7$ to $-5.8$) corresponds to the tetrahedral Ga sites of the $\gamma$-phase ($\gamma$-\ce{Ga_{Td}}) (shown in the inset), which occupy the cation sites of a perfect spinel structure. This feature serves as key evidence for the $\gamma$-phase transition. The maximum depths for the four surfaces \hkl(100), \hkl(010), \hkl(001) and \hkl(-201) are 150, 210, 210, and 180~\r A, respectively.}
    \label{fig:fig3}
\end{figure}

We further employed the local entropy method developed by Piaggi and Parrinello~\cite{2017Localentropy} for the identification and characterization of the individual \ce{Ga2O3} polymorphs ($\beta$, $\gamma$, and amorphous phases).
This approach computes a local entropy value for each atom based on its atom-centered RDF, generating a structural fingerprint that quantifies atomic-scale order-disorder characteristics~\cite{2017Localentropy}.
We validated the method by benchmarking it against the pristine $\beta$, $\gamma$, and amorphous phase models, which confirmed its effectiveness in distinguishing these phases.
The benchmarking procedure, along with the specific parameter settings and implementation details, is documented in Supplementary Note 2, Figures 12-14.
Fig.~\ref{fig:fig3} summarizes the structural evolution and depth-dependent distribution of the four irradiated surface models, as visualized by color-coded local entropy analysis.
Figs.~\ref{fig:fig3}a-d display the atomic configurations at a depth of 60~\r A for all four surfaces, color-coded by Ga-O local entropy (the complete depth-dependent snapshots are given in Supplementary Figures~15-18).
Based on the Ga-O local entropy, the amorphous, disorder $\gamma$, more ordered $\gamma$ and $\beta$ phases of \ce{Ga2O3} can be effectively distinguished.
The configurations in Regions I-V (identified in Figs.~\ref{fig:fig3}a-b and presented in Fig.~\ref{fig:fig3}e) confirm these entropy-defined phases.
Region I, with an entropy range of $-2.5$ to $-1.5$, exhibits a disordered atomic arrangement, confirming its amorphous nature.
Region II displays the characteristic atomic stacking of the $\gamma$-phase within its defined local entropy range ($-3.5$ to $-2.5$), however, it exhibits less structural order than Region IV.
Region III and IV on the \hkl(100) surface exhibit the atomic structure of the $\beta$- and $\gamma$-phases ($-4.5$ to $-3.5$), with the $\gamma$-phase adopting a square structure along the \hkl(100) orientation.
Region V illustrates the structural evolution on the $\beta$-\hkl(010) surface, which includes disordered $\gamma$-phase, more ordered $\gamma$-phase, and stable $\beta$-phase configurations.
In regions of minor radiation damage, point defects remain stable within the $\beta$-phase matrix~\cite{2025ActaTaiqiao} and are insufficient to drive a phase transition.
In contrast, once irradiation-induced defects accumulate to a degree that locally destabilizes the $\beta$-phase lattice, it promotes the nucleation of the $\gamma$-phase.
It can be seen that $\gamma$-\ce{Ga2O3} can be regarded as a defective $\beta$-\ce{Ga2O3}, as noted in prior work~\cite{2025scripta}.
The overall $\beta$-to-$\gamma$ phase transition is driven by the rapid recovery of the O sublattice and rearrangement of Ga atoms, yet the specific crystallographic orientation of the $\gamma$-phase is dictated by the atomic anisotropy of the initial $\beta$ surface.
This establishes a structural selection rule where the $\gamma$-phase develops along orientations that maximize structural similarity to the parent $\beta$ surface:
The $\beta$-\hkl(100) surface transforms into $\gamma$-\hkl(100) due to their inherent structural resemblance.
The strong channeling effects in $\beta$-\hkl(010) and the open structure of $\beta$-\hkl(001) each facilitate Ga interstitial rearrangement to form $\gamma$-\hkl(110), consistent with previous transmission electron microscopy (TEM) observations~\cite{gamma110APLMater} and X-ray diffraction (XRD) results~\cite{EpitaxialRecovery2025}.
The $\beta$-\hkl(-201) surface evolves into the $\gamma$-\hkl(111) orientation, with both sharing close structural similarity as to be nearly indistinguishable, an orientation relationship also been experimentally confirmed recently~\cite{bektas2025defect}.
The color-coded snapshots of the coexisting $\beta$- and $\gamma$-phases on the four surface orientations are presented in Supplementary Figures~19 and 20.

Although the local entropy of the Ga-O pair allows for effective visualization and distinguishes between amorphous and crystalline $\beta$/$\gamma$ phases, it fails to differentiate between the $\beta$ and $\gamma$ phases due to significant overlap.
As shown in Fig.~\ref{fig:fig1}, the primary difference between these phases lies in the arrangement of Ga sublattices.
Consequently, the radiation-induced changes in the probability density distribution based on Ga-Ga local entropy were calculated by subtracting the pristine $\beta$ profile from that of the irradiated structures.
Figs.~\ref{fig:fig3}f-i present the local entropy difference for the irradiated \hkl(100), \hkl(010), \hkl(001) and \hkl(-201) surfaces, sampled at depths of 15, 60, and 105~\r A, as well as at the maximum model depth (the full depth-dependent evolution of local entropy across each surface is provided in Supplementary Figure~21).
Consistent with the benchmark established in Supplementary Figure~14, the amorphous phase is predominantly distributed in the region above $-3.6$.
The intermediate range ($-5.8$ to $-3.6$) corresponds to the depletion of $\beta$-phase and the emergence of the $\gamma$-phase.
Crucially, the region below $-5.8$ (highlighted by the red shaded box) reveals distinct signatures of the $\gamma$-phase.
As illustrated in the inset of Fig.~\ref{fig:fig3}f, the specific entropy range corresponds to the tetrahedral Ga sites of $\gamma$-phase ($\gamma$-\ce{Ga_{Td}}), which structurally coincide with the cation sites of a perfect spinel.
The extent of amorphization gradually diminishes with increasing depth.
In the near-surface region within 60~\r A, the amorphous phase and the $\gamma$-phase exhibit comparable formation probabilities.
Notably, the fraction of $\gamma$-\ce{Ga_{Td}} displays a non-monotonic trend that initially increases and subsequently decreases with depth, indicating that the $\beta$-to-$\gamma$ phase transition is predominantly localized near the surface.
The \hkl(010) surface retains the highest fraction of $\gamma$-\ce{Ga_{Td}} and the lowest degree of amorphization.
Conversely, the \hkl(-201) surface shows the opposite behavior, characterized by the most severe amorphization and suppressed $\gamma$-\ce{Ga_{Td}} formation.
The irradiated \hkl(100) and \hkl(001) surfaces yield comparable quantities of $\gamma$-\ce{Ga_{Td}}.
Overall, this depth- and orientation-resolved analysis demonstrates that surface-specific atomic arrangements govern both the extent and distribution of radiation-induced structural disorder, underscoring the decisive role of crystallographic orientation in radiation tolerance.

\begin{figure}[ht!]
    \centering
    \includegraphics[width=16cm]{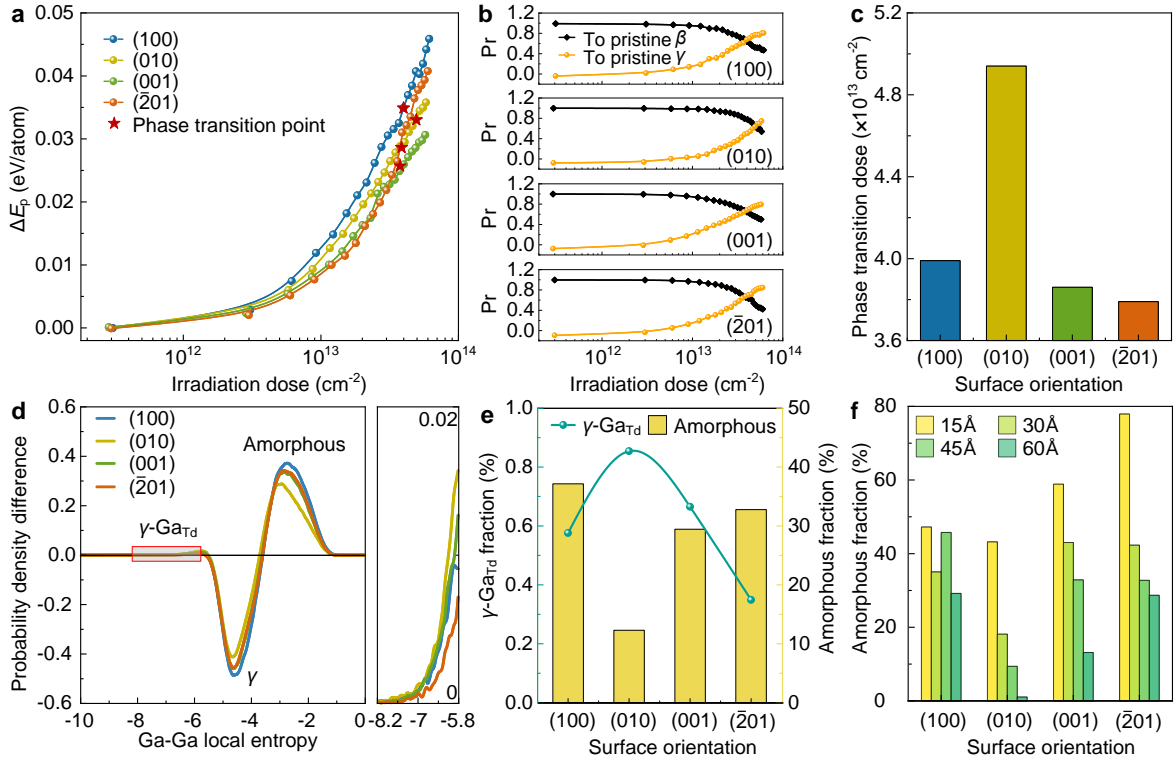}
    \caption{\textbf{Dose-dependent structural pathways in the $\beta$-phase and the role of surface orientation.} 
    \textbf{a} $\Delta E_\mathrm{p}$ relative to the pristine $\beta$-phase as a function of dose.
    \textbf{b} Evolution of the Pr correlation coefficient with increasing irradiation dose.
    \textbf{c} Comparison of phase transition doses across different surface orientations. 
    \textbf{d} Local entropy difference for each surface at its respective critical dose in \textbf{c}. The magnified region outlined in red (range: $-8.2$ to $-5.8$) corresponds to $\gamma$-\ce{Ga_{Td}}.
    \textbf{e} $\gamma$-\ce{Ga_{Td}} and amorphous frictions at the depth range 20--60~\r A.
    \textbf{f} Amorphous phase fraction for each 15-\r A layer within the 60~\r A.}
    \label{fig:fig4}
\end{figure}

To examine the dose-dependent evolution of the phase transition, the potential energy change ($\Delta E_\mathrm{p}$) relative to the pristine $\beta$-phase was calculated.
$\Delta E_\mathrm{p}$ increases monotonically with irradiation dose without reaching saturation (Fig.~\ref{fig:fig4}a), indicating that the phase transition remains ongoing throughout the irradiation process.
To minimize the effect of surface amorphization, we focused on the 20--60~\r A region below the surface where the $\gamma$-phase transition takes place.
Within this layer, the structural similarity to the $\gamma$-phase increases with irradiation dose across all surfaces, accompanied by a corresponding decrease in similarity to the $\beta$-phase (Fig.~\ref{fig:fig4}b).
After irradiation, the $\gamma$-phase similarities reach 0.81 for \hkl(100), 0.75 for \hkl(010), 0.79 for \hkl(001), and 0.85 for \hkl(-201), revealing orientation-dependent responses.
The critical dose required to initiate the $\gamma$-phase transition varies substantially with surface orientation (Fig.~\ref{fig:fig4}c).
The \hkl(010) surface is the most radiation-resistant, requiring the highest dose for transition.
The critical dose then decreases through the series \hkl(100), \hkl(001), to \hkl(-201).
This trend confirms the superior radiation tolerance of the \hkl(010) surface, consistent with our previously reported channeling effect that effectively mitigates defect accumulation~\cite{2025ActaTaiqiao}.
The local entropy difference derived from the Ga-Ga pair within the 20--60~\r A depth range reveals distinct structural evolution across surfaces (Fig.~\ref{fig:fig4}d, corresponding structural mappings provided in Supplementary Figure~22).
Although all surfaces undergo amorphization in this region, the extent varies significantly.
The \hkl(010) surface exhibits the greatest resistance to amorphization (Fig.~\ref{fig:fig4}e) and concurrently retains the highest proportion of $\gamma$-\ce{Ga_{Td}}, indicating a preferential pathway toward $\gamma$-phase formation.
In contrast, the \hkl(-201) surface, while showing slightly less amorphization than the \hkl(100) surface at the given dose, forms the least amount of $\gamma$-\ce{Ga_{Td}}.
Notably, the amorphous fraction peaks at a shallower depth of 15~\r A (Fig.~\ref{fig:fig4}f), reaching approximately $80\%$ for the \hkl(-201) surface.
These results underscore a competitive process between amorphization and $\gamma$-phase transition, with the extent of $\gamma$-\ce{Ga_{Td}} formation serving as a direct indicator of the latter.
The pronounced differences provide clear evidence for a significant $\gamma$-phase transition at the applied irradiation dose.

\begin{figure}[ht!]
    \centering
    \includegraphics[width=16cm]{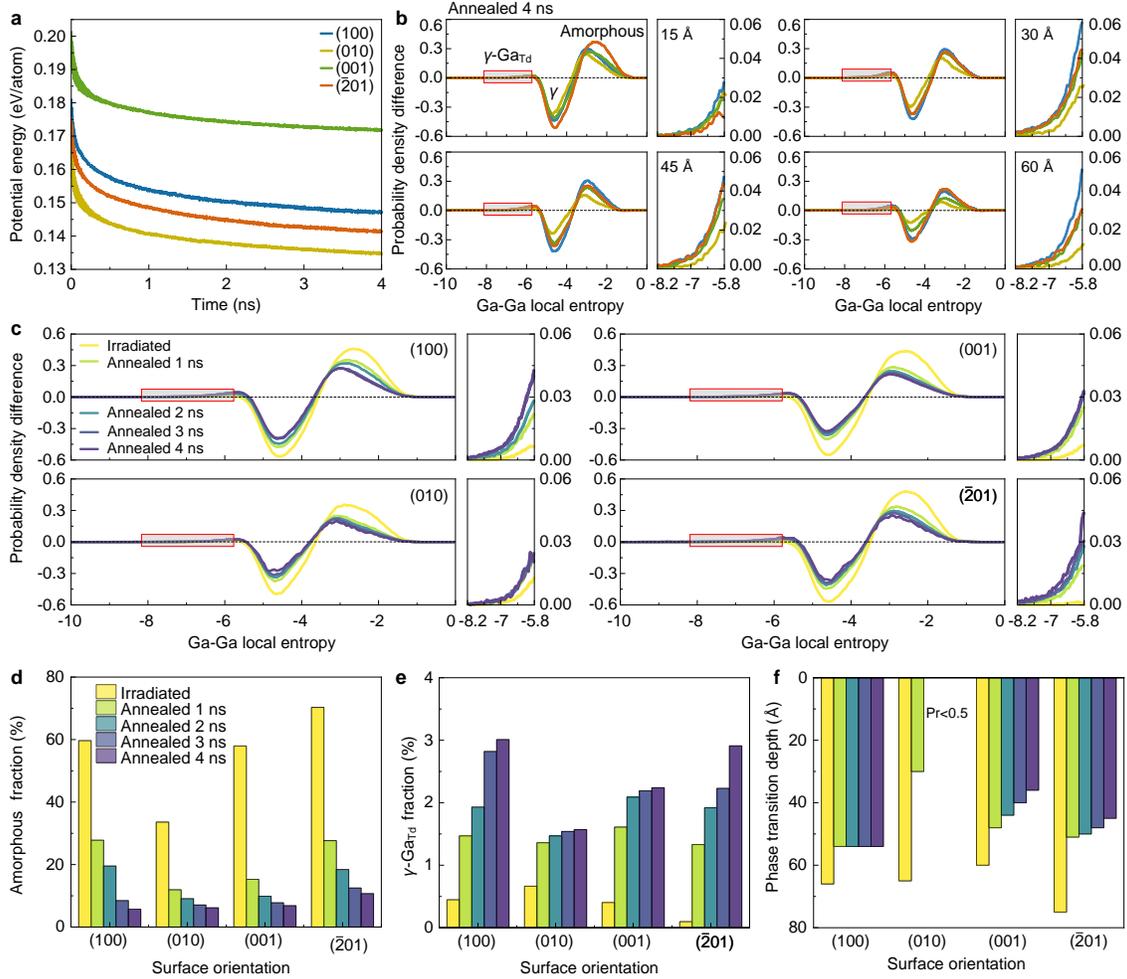}
    \caption{\textbf{Phase evolution and structural recovery during thermal annealing for 4~ns at 1200~K.} 
    \textbf{a} $\Delta E_\mathrm{p}$ relative to the pristine $\beta$-phase as a function of annealing time. 
    \textbf{b} Depth-dependent distribution of Ga-Ga local entropy difference after 4~ns of annealing.
    \textbf{c} Evolution of the Ga-Ga local entropy probability density difference with annealing time (1--4~ns), shown for all four surfaces over depths of 0-60~\r A.
    \textbf{d,e} Depth-integrated (0--60~\r A) phase fractions quantified from Ga-Ga local entropy difference, showing the reduction in amorphous content (d) and the redistribution of $\gamma$-\ce{Ga_{Td}} populations (e) after annealing. The magnified region outlined in red (range: $-8.2$ to $-5.8$) corresponds to $\gamma$-\ce{Ga_{Td}}.
    \textbf{f} Phase transition-layer depth for each surface, determined from the crossover point of Pr similarity in Supplementary Figure~28.}
    \label{fig:fig5}
\end{figure}

To assess the stability of the $\gamma$ and amorphous phases, the systems were annealed at 1200 K for 4~ns.
Fig.~\ref{fig:fig5} tracks the structural recovery and phase evolution in $\beta$-\ce{Ga2O3} during 4~ns of post-irradiation annealing.
The continuous decrease in system potential energy ($\Delta E_\mathrm{p}$) relative to the pristine $\beta$-phase (Fig.~\ref{fig:fig5}a) demonstrates efficient damage recovery and structural ordering.
Ga-Ga local entropy difference in Fig.~\ref{fig:fig5}b reveals distinct depth-dependent recovery patterns (for atomic snapshots based on Ga-O local entropy see Supplementary Figures~15-18; for local entropy distributions based on Ga-Ga local entropy see Supplementary Figures~23-26).
The \hkl(-201) surface remains the highest amorphous fraction at 15~\r A whereas the other surfaces remain largely crystalline.
Compared to the irradiated state in Fig.~\ref{fig:fig3}, both the amorphous and $\gamma$-phase fractions decrease after annealing.
This reduction is particularly pronounced in the region around approximately 60~\r A, which was predominantly $\gamma$-phase after irradiation but shows a substantially lower $\gamma$-phase proportion after 4~ns of annealing.
The signature of $\gamma$-phase transition $\gamma$-\ce{Ga_{Td}} is primarily concentrated between 15--45~\r A.
Even at the shallow depth of 15~\r A, the $\gamma$-\ce{Ga_{Td}} fraction is more than that in the irradiated state, suggesting that annealing promotes the transition of disordered $\gamma$-phase and amorphous regions into a more ordered $\gamma$-phase structure.
Combined analysis of the Ga-Ga local entropy difference (Fig.~\ref{fig:fig5}c) and the amorphous fraction derived from Ga-O analysis (Fig.~\ref{fig:fig5}d) shows that the amorphous regions have largely recovered, demonstrating the excellent self-healing capability of \ce{Ga2O3}.
The quantitative orientation- and depth-dependent distributions of $\gamma$-\ce{Ga_{Td}} and the amorphous phase can be seen in Supplementary Figure~27.

Depth-integrated analysis of the 0--60~\r A near-surface region reveals pronounced crystallographic anisotropy in the kinetic stability of the $\gamma$-phase (Figs.~\ref{fig:fig5}d-f).
The degree of amorphization after irradiation follows the order: \hkl(010) $<$ \hkl(001) $<$ \hkl(100) $<$ \hkl(-201) (Fig.~\ref{fig:fig5}d).
After annealing, all surfaces except \hkl(-201) converge to similarly low amorphous fractions, with \hkl(-201) retaining the highest residual disorder.
However, the recovery pathways diverge fundamentally.
For the non-channeling surfaces \hkl(100), \hkl(001), \hkl(-201), the structural recovery proceeds via a two-step recrystallization mechanism.
Following irradiation-induced amorphization and limited $\gamma$-phase formation, annealing first drives the transition of the amorphous phase into the $\gamma$-phase, which then converts into the thermodynamically stable $\beta$-phase.
This sequential amorphous-to-$\gamma$-to-$\beta$ pathway accounts for the accumulation of the $\gamma$-phase observed near the surface during recovery.
The \hkl(010) surface exhibits a distinct and direct pathway.
At both specific depths (Fig.~\ref{fig:fig5}b) and over the integrated 0-60~\r A region (Fig.~\ref{fig:fig5}c,e), it shows the strongest $\gamma$-\ce{Ga_{Td}} signal after irradiation among all surfaces, indicating a high propensity for $\gamma$-phase formation under irradiation with limited amorphization.
After annealing, it retains the lowest $\gamma$-phase fraction, suggesting that the $\gamma$-phase formed on this surface is relatively unstable.
Therefore, in addition to following the general recovery pathway, the \hkl(010) surface likely undergoes a more direct recovery mechanism where a significant portion of the irradiation-induced $\gamma$-phase converts directly to the $\beta$-phase, indicating a reversible recovery process.
The evolution of $\gamma$-phase transition depth in Fig.~\ref{fig:fig5}f further confirms this distinction.
The depth was determined from the intersection of Pr correlation coefficients (complete Pr profiles in Supplementary Figure~28).
During annealing, the phase transition depth of the \hkl(100) surface remains stable near 54~\r A, demonstrating that its $\gamma$-phase is kinetically stable at this depth.
For the \hkl(010) surface, the Pr curves show no intersection and similarity falls below 0.5 after annealing for more than 3~ns, indicating minimal $\gamma$-phase fraction as it has largely reverted to the $\beta$-phase.
Annealing of ion-implanted samples (fluence: $10^{14}$~cm$^{-2}$) shows that the recovery of the $\gamma$-phase begins at 773~K and finishes at 1173~K~\cite{EpitaxialRecovery2025}.
The strong consistency between our MD simulation of ion irradiation (fluence: $10^{13}$~cm$^{-2}$, annealing at 1200~K) and existing experimental data underscores the robustness of our findings.
On \hkl(001) and \hkl(-201) surfaces, the transition depth progressively recedes toward the surface, indicating lower kinetic stability compared to \hkl(100).
Our atomic-level description establishes a mechanistic foundation for the macroscopic phenomena reported in the literature~\cite{2023radiation, 2025NanoLett, 2025PhysRevLett, 2025scripta, EpitaxialRecovery2025, bektas2025defect}.
These quantitative results establish that the strong anisotropy in $\gamma$-phase stability roots in the crystallographic-orientation-dependent recovery pathway, which directly dictates the kinetics of its nucleation and depth-dependent evolution.

In conclusion, our study uncovers a novel paradigm for radiation damage in functional oxides, revealing that the irradiation response of $\beta$-\ce{Ga2O3} is governed not by stochastic amorphization but by a crystallographically anisotropic, phase-transition-mediated pathway.
By combining ML-MD simulations with a local entropy-based structural descriptor, we quantitatively resolve the anisotropic kinetics of the $\beta$-to-$\gamma$ transition.
We establish that the critical dose, transition-layer depth, and long-term phase stability are fundamentally governed by surface orientation.
The irradiation-induced $\beta$ to $\gamma$ phase transition exhibits distinct crystallographic orientation relationships for each surface: the $\beta$-\hkl(100) surface forms $\gamma$-\hkl(100); the $\beta$-\hkl(010) and \hkl(001) surfaces produce $\gamma$-\hkl(110); and the $\beta$-\hkl(-201) surface yields $\gamma$-\hkl(111).
The $\gamma$-phase on the \hkl(100) surface maintains a transition-layer depth near 54~\r A, whereas on the \hkl(010), \hkl(001), and \hkl(-201) surfaces, the $\gamma$-phase progressively recedes toward the surface upon annealing.
Crucially, these orientation relationships, together with divergent recovery behaviors during annealing, reveal two competing transition pathways.
The \hkl(010) surface undergoes direct $\gamma$-to-$\beta$ reversion, whereas non-channeling surfaces follow a sequential amorphous-to-$\gamma$-to-$\beta$ pathway. 
These insights not only overturn the conventional amorphization-centric view of radiation damage but also establish surface selection as a general design principle for achieving predictable radiation tolerance in functional oxides.
The methodology and mechanistic understanding developed here provide a universal framework for designing materials with enhanced performance in extreme irradiation environments.

\newpage
\subsection*{Methods}

\textbf{MD simulations.}
All machine-learning molecular dynamics (ML-MD) simulations were performed using the large-scale atomic/molecular massively parallel simulator (LAMMPS) code~\cite{2022la}.
Ion irradiation simulations utilized a machine-learning potential trained on ab initio data and extensively validated for \ce{Ga2O3}, which accurately models the polymorphs and short-range repulsive interactions~\cite{Zhaonpj2023, 2023radiation, 2025ActaTaiqiao, 2025NanoLett, 2025PhysRevLett, 2025alphaNC, he2024threshold}.
The surface orientations of $\beta$-\ce{Ga2O3} investigated for damage accumulation were the \hkl(100), \hkl(010), \hkl(001) and \hkl(-201), as illustrated in Fig.~\ref{fig:fig1}.
Initial thermalization of the bulk cell was conducted at 0~bar and 300~K using the isothermal-isobaric ($NPT$) ensemble for 50~ps.
The simulation cells were designed to be sufficiently large to ensure reliable results, with dimensions and atom counts as follows: $\sim 177 \times 186 \times 230$~\r A$^{3}$ (504,000 atoms) for \hkl(100), $\sim 194 \times 177 \times 295$~\r A$^{3}$  (729,600 atoms) for \hkl(010), $\sim 188 \times 186 \times 290$~\r A$^{3}$ (720,000 atoms) for \hkl(001), $\sim 186 \times 180 \times 254$~\r A$^{3}$ (576,000 atoms) for \hkl(-201), respectively.
Each model included a 40~\r A vacuum layer above the surface and a 20~\r A vacuum layer at the bottom.
Irradiation simulations used 5~keV Ga ions with 200 cascade events at 300~K.
To prevent channeling effects, Ga ion was set at a $7\degree$ incidence angle relative to the surface normal~\cite{Nordlund-PhysRevB.94.214109}.
An adaptive MD timestep ensured the maximum atomic displacement per step remained below 0.1~\r A in cascade regions.
Electron stopping is applied as a friction term to atoms with kinetic energy greater than 10~eV~\cite{PRB1998nordlund}.
Irradiation simulations were conducted using 5~keV energy at 300~K, with a total of 200 cascade collision events.
Each cascade was run for 30~ps in the microcanonical ($NVE$) ensemble.
Following each cascade event, the entire structure was randomly translated along the in-plane directions to ensure stochastic impact locations.
Prior to the subsequent cascade, the system was subjected to a 20 ps relaxation in the canonical ($NVT$) ensemble using Nos{\'e}-Hoover thermostat~\cite{Hoover1985PRA} to equilibrate the structure.
The \hkl(100), \hkl(010), \hkl(001) and \hkl(-201) surfaces were irradiated to doses of 6.14/5.81/5.74/5.90 $\times 10^{13}$~cm$^{-2}$, respectively, following the described simulation procedure.
Each irradiated surface was then subjected to a 4~ns annealing at 1200~K.

\textbf{Local entropy fingerprint for atomic environment identification.}
The local entropy can distinguish extremely well between solid-like and liquid-like atoms, a capability that remains effective even in inhomogeneous situations where different atomic environments coexist.
The key innovation lies in projecting this global entropy onto individual atoms to create local fingerprints.
The atom-projected local entropy is defined as~\cite{2017Localentropy}:
\begin{equation}
s_S^i = -2\pi\rho k_B\int_0^{r_m} \left[ g_m^i(r) \ln g_m^i(r) - g_m^i(r) + 1 \right] r^2  dr,
\end{equation}
where $\rho$ is the system's density,  $k_B$ is the Boltzmann constant, and $g(r)$ is the radial distribution function. $r_{m}$ is an integration cutoff that ensures locality while capturing essential structural information, and $g_{m}^{i}(r)$ represents a mollified radial distribution function centered on atom $i$. 
The method has been fully integrated into the Open Visualization Tool (OVITO) package~\cite{2017Localentropy, 2010ovito}, with implementation details provided in Supplementary Note~2.

\subsection*{Data availability}

The data that support the findings of this study are available within the paper and its Supplementary Information file. 
The corresponding raw data of the classical MD simulations published in this paper are openly available at \url{https://doi.org/10.6084/m9.figshare.xxxxxxxx}.

\subsection*{Code availability}

The code and software used in this work are LAMMPS and OVITO, which are openly available online from the corresponding developers and maintainers.

\subsection*{Acknowledgments}

The project was supported by the Major Program (JD) of Hubei Province (Grant No. 2023BAA009), the National Natural Science Foundation of China (Grant Nos. 52302046,  T252790007, T2525025, and L2424216). We also thank the Supercomputing Center of Wuhan University for their support of the calculation.

\subsection*{Author contributions}

Z.Z. and J.Z. conceived the research strategy and supervised the project. T.L. performed all simulations, conducted the data analysis, and wrote the initial draft of the manuscript with advice from Z.Z., J.Z., Y.G., F.D., and S.L. T.W., E.Z., J.F., and X.F. assisted with data analysis and figure preparation. Z.L. contributed the analytical methodological concept. F.D. helped validate the analytical approach. J.Z. and Z.Z guided the development of the analytical methodology and framed the data interpretation. Z.Z. conceptualized this project framework and provided the funding acquisition. All authors discussed the results and approved the final manuscript.

\subsection*{Competing interests}

The authors declare no competing financial interests.

\subsection*{Supplementary information}
The online version contains
supplementary material available at \url{https://doi.org/xxxxxxxx}

\bibliography{ref}

\end{document}